\documentclass[checkin,prb]{revtex4-1}

\newcommand{\br}{{\bf r}}

\usepackage{amsmath}
\usepackage[all]{xy}
\usepackage[T1]{fontenc}
\usepackage[latin9]{inputenc}

\newcommand{\clr}{}%{\color{red}}%{\color{yellow}}%{\color{black}}

\newcommand{\mje}{mJ/cm$^2$}

\newcommand{\bp}{{\bf p}}
\newcommand{\bs}{{\bf s}}

\newcommand{\bn}{\begin{enumerate}}
\newcommand{\en}{\end{enumerate}}
\newcommand{\ba}{\begin{eqnarray}}
\newcommand{\ea}{\end{eqnarray}}

\usepackage{graphicx,color}

\newcommand{\ib}{\frac{2\pi}{a}}

\newcommand{\R}{\rightarrow}
\newcommand{\chl}[4]{$|#1\ra\R|#2\ra\R|#3\ra\R|#4\ra$}
\newcommand{\ket}[1]{$|#1\ra$}

\newcommand{\be}{\begin{equation}}
\newcommand{\ee}{\end{equation}}
\newcommand{\la}{\langle}

\newcommand{\ra}{\rangle}
\newcommand{\et}{{\it et al. }}

\newcommand{\nk}{{n{\bf k}}}

\newcommand{\bk}{{\bf k}}

\def\prl{{ Phys. Rev. Lett. }}

\newcommand{\mn}{Mn$_2$RuGa }
\newcommand{\mne}{Mn$_2$RuGa}
\newcommand{\mrg}{Mn$_2$RuGa }

\begin{document}

%\title{ Theory of ultrafast demagnetization:\\
%Perspectives from  spin-orbit-coupled Heisenberg system}

%\title{Demagnetization in a spin-orbit-coupled Heisenberg
%  system:\\ Applications to ultrafast spin dynamics}

%\title{Simple picture of demagnetization from a spin-orbit-coupled
%  Heisenberg system: From static to dynamic}

%\title{Essence of femtosecond demagnetization:\\ Exact results from a
%  spin-orbit-coupled Heisenberg system}

%\title{Understanding femtosecond magnetism:\\ A simple picture
%  from a spin-orbit-coupled Heisenberg system}

%\title{Exchange interaction  in femtomagnetism}
%\title{Electron exchange-interaction collapse as an alternative
%  mechanism for femtomagnetism }

%\title{A path to the consistent theory of femtosecond
%  magnetism:\\ Spin-orbit-coupled Heisenberg exchange model
%}

%\title{Prediction of lasing in Mn$_2$RuGa away from the Weyl points}

%\title{Electron momentum-enabled group-velocity decomposition and
%  X-ray spectroscopy}

%\title{Lissajous orbits of momentum on an ultrafast time scale}

%\title{Emergence of the high-order optical Pancharatnam-Berry tensor
%  from all-optical spin switching in ferrimagnetic Heuslers}

\title{Gateway to all-optical spin switching in % noncentral symmetric
  Heusler ferrimagnets:  Pancharatnam-Berry tensor  and magnetic
  moment ratio
}

\author{G. P. Zhang$^*$} \affiliation{Department of Physics, Indiana
  State University, Terre Haute, IN 47809}

\author{Y. Q. Liu} \affiliation{School of Materials and Energy,
  Lanzhou University, Lanzhou 730000, China}

\author{M. S. Si} \affiliation{School of Materials and Energy, Lanzhou
  University, Lanzhou 730000, China}

\author{Nicholas Allbritton} \affiliation{Department
  of Physics, Indiana State University, Terre Haute, IN 47809,
  USA}

%\author{Robert Meadows} \affiliation{Department
%  of Physics, Indiana State University, Terre Haute, IN 47809,
%  USA}

%\author{Antonio Tamayo} \affiliation{Department
%  of Physics, Indiana State University, Terre Haute, IN 47809,
%  USA}

\author{Y. H. Bai} \affiliation{Office of Information
  Technology, Indiana State University, Terre Haute, Indiana 47809,
  USA}

 \author{Wolfgang H\"ubner} \affiliation{Department of Physics,
   Rheinland-Pf\"alzische Technische Universit\"at
   Kaiserslautern-Landau, 
   67653 Kaiserslautern, Germany}

 \author{Thomas F. George} \affiliation{Departments of Chemistry \&
   Biochemistry and Physics \& Astronomy \\University of
   Missouri-St. Louis, St.  Louis, MO 63121 }

%     to correlate the spin dynamics to the
%    second-order optical response and then

\date{\today}

\begin{abstract}

{All-optical spin switching (AOS) is a new phenomenon found in a small
  group of magnetic media, where a single laser pulse can switch spins
  from one direction to another, without assistance of a magnetic
  field, on a time scale much shorter than existing magnetic
  technology. However, despite intensive efforts over a decade, its
  underlying working principle remains elusive. Here through
  manganese-based Heusler ferrimagnets, we show that a group of flat
  bands around the Fermi level act as gateway states to form efficient
  channels for spin switching, where their noncentrosymmetry allows us
  to correlate the spin dynamics to the second-order optical response.
  To quantify their efficacy, we introduce the third-rank
  Pancharatnam-Berry tensor (PB tensor), $\boldsymbol{\eta}^{(3)}=\la
  i |\bp |m\ra \la m|\bp |f\ra \la f|\bp |i\ra,$ where $|i\ra$,
  $|m\ra$ and $|f\ra$ are {\clr initial, intermediate and final band
    states, respectively}, and $\bp$ is the momentum operator.  A
  picture emerges: Those which show AOS, such as the recently
  discovered \mne, always have {\clr a large PB tensor element} but
  have a small sublattice spin moment ratio, consistent with the prior
  experimental small remanence criterion.  This does not only reveal
  that the delicate balance between the large PB tensor element and
  the small sublattice spin ratio plays a decisive role in AOS, but
  also, conceptually, connects the $n$th-order nonlinear optics to
  $(n+1)$th-rank PB tensors in general.  }
\end{abstract}

%\pacs{75.78.Jp, 75.40.Gb, 78.20.Ls, 75.70.-i}

%ultrafast magnetization dynamics, 75.78.Jp

%75.40.Gb, 78.20.Ls, 75.70.-i, 78.47.J-}

%\keywords{femtomagnetism; exchange interaction}

\maketitle

 \section{Introduction}

%\linenumbers
%\lipsum[1-5]

  For over half a century, magnetic storage technology has relied on a
magnetic field to switch magnetization in magnetic drives.
Beaurepaire and coworkers \cite{eric} changed this picture by showing
that a femtosecond laser pulse can directly demagnetize the magnetic
nickel film within 1 ps, representing the beginning of femtomagnetism
\cite{ourreview}.  In 2007, Stanciu and coworkers \cite{stanciu2007}
reported that a single laser pulse alone, without assistance of a
magnetic field, is able to switch spins from one direction to another
in GdFeCo, called all-optical spin switching (AOS).  Subsequent
investigations
\cite{ostler2012,mangin2014,lambert2014,mplb18,quessab2019,lin2023,wang2023,kasatani2023}
soon discovered a large group of AOS materials
\cite{alebrand2012,hassdenteufel2013,mangin2014,lambert2014,vomir2017}. It
has been shown that whether or not AOS occurs sensitively depends on
the sample composition \cite{vahaplar2012,hassdenteufel2013},
thickness \cite{hebler2016,gorchon2017}, remanence
\cite{hassdenteufel2015}, number of laser pulses, laser fluences and
other parameters \cite{vahaplar2012} (see recent reviews for details
\cite{mplb18,scheid2022}).  Mentink \et \cite{mentink2012} suggested
an intersite spin exchange model for AOS.  Recently Jakobs and Atxitia
\cite{jakobs2022b} proposed a criterion based on a transition from a
relativistic relaxation path to an exchange-dominated regime.

 Unfortunately, most AOS materials are amorphous, which significantly
 limits access to their microscopic electronic and magnetic
 structures, beyond the classical atomistic spin model level
 \cite{evans2014,vahaplar2012}.  Although crystalline AOS materials
 are also discovered \cite{vomir2017,parlak2018,kichin2019}, they
 often need multiple laser pulses to switch spins
 \cite{alebrand2012,alebrand2012a}, thus not ideal for future
 applications.  Heusler alloys are different
 \cite{wurmehl2006,graf2009,endo2012,hakimi2013,wollmann2014,yang2015,zic2016,chang2016,wang2016,kubler2016,song2017,zhang2018a,ernst2019,prb22}. They
 are crystalline and more importantly their electronic and magnetic
 properties can be systematically tuned \cite{wollmann2017} without a
 significant change to their structures.  What is even more
 interesting is that one of the Heusler compounds, \mne, shows
 single-pulse all-optical spin switching \cite{banerjee2020}.  This
 presents a rare opportunity for a first-principles investigation,
 beyond earlier model simulations
 \cite{aip20,jakobs2022b,jakobs2022c}.

In this {\it Letter}, we employ four manganese-based
noncentrosymmetric Heuslers $\rm Mn_2YZ$ and the time-dependent
first-principles method to pin down two critical elements that
underline the all-optical spin switching.  First, a group of flat
bands, a few eV around the Fermi level, act as gateway states for spin
switching.  To quantify the efficacy of these states, we introduce the
Pancharatnam-Berry (PB) tensor, a geometric tensor that consists of a
product of momentum matrix elements between transition
states. Channels involving flat bands have a large PB tensor element, found
across all the four Heuslers.  Second, the spin moment ratio between
two Mn sublattices must be small. This is only found in \mn and $\rm
Mn_2RuAl$, not in $\rm Mn_2IrGa$ and $\rm Mn_2RuGe$.  Our finding
conceptually establishes the link between the $(n+1)$th PB tensor to
the $n$th order optical process, and reveals that it is the balance
between {\clr a large PB tensor element} and a small spin moment ratio that
leads to all-optical spin switching. Our finding will have a profound
impact on the future direction of the experimental and theoretical
research, by focusing on crystalline Heuslers. In particular, since
all four materials investigated here are already available, this will
motivate an immediate experimental investigation.

%\section{Correlation between the spin and charge dynamics}  

\section{Correlation between the spin and charge dynamics and
  Pancharatnam-Berry momentum tensor}

AOS is an optical process.
What is probed often differs from what one wishes to probe, similar to
time-resolved magneto-optics \cite{np09}.  Figure \ref{fig1}(a)
schematically illustrates a laser pulse impinging on a sample, where
one detects the polarization change and derives the spin change from
it. Theoretically, {\clr we first solve the Kohn-Sham equation for the
  ground state properties \cite{wien2k},
\be \left
[-\frac{\hbar^2\nabla^2}{2m_e}+V_{Ne}+V_{H}+V_{xc} \right
]\psi_{\nk}(\br)=E_{\nk} \psi_{\nk} (\br), \label{ks} \ee where
$\psi_{\nk}(\br)$ is the wavefunction of band $n$ at the crystal
momentum ${\bf k}$, and $E_{\nk}$ is its band energy.  The terms on the
left are the kinetic energy operator, the attraction between the
nuclei and electrons, the Hartree term, and the exchange-correlation,
respectively.  The spin-orbit coupling (SOC) is included using a
second-variational method in the same self-consistent iteration.
We describe the optical
  interaction with a matter by the time-dependent Liouville equation
  \cite{np09,jpcm23} (additional details can be found in the
  supplementary material (SM) \cite{sm})}, \be \frac{d \rho}{d
  t}=\frac{1}{i\hbar}[H_0+H_I, \rho],
\label{liu1} \ee
 where $\rho$ is the density matrix, $H_0$ is the unperturbed system
 Hamiltonian and $H_I$ is the interaction between the laser field and
 the system \cite{sm}. {\clr  The  Liouville equation
   builds in the Pauli exclusion principle, can be extended to
   many-body states,  and
   works well with other softwares such as VASP, Quantum Espresso, and
   ELK}.  
 The $n$th-order density matrix is given by \cite{butcher}
\begin{widetext}
 \be
 \rho^{(n)}(t)=(i\hbar)^{-n}U_0(t)\int_{-\infty}^tdt_1
 \int_{-\infty}^{t_1} dt_2\cdots \int_{-\infty}^{t_{n-1}}
 dt_n[H_I(t_1),[H_I(t_2),\cdots[H_I(t_n),\rho_0]\cdots ]] U_0(-t), \ee
\end{widetext}
 where $U_0(t)=\exp(-iH_0t/\hbar)$.  The symmetry property of electron
 and spin dynamics is encoded in the expectation value of an
 operator. For the reason that will become clear below, we take the
 second-order momentum $\la \bp \ra^{(2)}(t)={\rm Tr}[\bp
   \rho^{(2)}(t)]$ as an example.  Under a symmetry operation $\{R\}$,
 $\la \bp \ra^{(2)}(t)$ transforms as a tensor (in the component form)
 \be \la p \ra^{(2)}_{lmn}(t)=\sum_{abc} R_{la}R_{mb}R_{nc} \la p
 \ra^{(2)}_{abc}(t), \ee where $l$ and $a$ are the new and old
 Cartesian indices of $\bp$, respectively, and the index pairs $(mb)$
 and $(nc)$ are Cartesian indices of two laser fields before and after
 symmetry transformation.  The entire procedure is exactly the same as the
 transformation between two susceptibility tensors $\chi^{(2)}$ and
 $\chi^{(2)'}$ in nonlinear optics as \cite{butcher} %\be
 $\chi^{(2)'}_{lmn}=\sum_{abc}R_{la}R_{m b} R_{nc} \chi^{(2)}_{abc}.$
% \ee

The situation of spin is different. Spin is an axial vector and
transforms under the SU(2) symmetry \cite{birss,nc18}.  For the same
second-order response, the spin expectation value, $\la \bs
\ra^{(2)}(t)={\rm Tr}[\bs \rho^{(2)}(t)]$, transforms as in a similar
fashion as the Hall effect \cite{lax} \be \la s
\ra^{(2)}_{lmn}(t)=\sum_{abc} R_{la}R_{mb}R_{nc}^s \la s
\ra^{(2)}_{abc}(t), \label{spin} \ee where the transformation matrix
for spin $R_{nc}^s$ is given as $R_{nc}^s=|R|R_{nc}$ and $|R|$ is its
determinant \cite{birss}. The second-order response in Eq. \ref{spin}
is the lowest-order response of the spin, and the first order is zero
because the product of one axial and one polar vector is zero
\cite{birss}, except that the axial and polar vectors are collinear
and the symmetry is low.  Because $\rho^{(2)}(t)$ is proportional to
the square of the laser field, that is why the (de)magnetization is
linearly proportional to the laser fluence as observed in experiments
\cite{regensburger2000,scheid2023} before saturation, and it is also
the reason why we choose the second-order density matrix above.
Regardless of whether the system has inversion symmetry or not, all
even orders of the spin, including the zeroth order, survive. However,
for the momentum, if the system has inversion symmetry, the
multiplication of three $R$'s yields $(-1)(-1)(-1)=-1$ and it cancels
out all even orders, including the second order here. Only odd orders
survive.  Therefore, in order to detect spin dynamics through the
charge dynamics or more precisely the optical response, at minimum one
must choose a system without inversion symmetry, so the second-order
$\la \bp^{(2)}\ra$ and $\la \bs^{(2)}\ra$ share the same field
dependence such as $H_I(t_1) H_I(t_2)$. {\clr If a system has inversion
symmetry, then there is a mismatch between spin and charge responses. 
 For a transition
 through a channel}
\ket{i}$\rightarrow$\ket{m}$\rightarrow$\ket{f}$\rightarrow$\ket{i},
the expectation value of the momentum $\la \bp\ra^{(2)}(t)$ is related
to the tensor \be \boldsymbol{\eta}^{(3)}(C)=\la i |\bp |m\ra \la
m|\bp |f\ra \la f|\bp |i\ra, \ee which is very similar to the
Pancharatnam-Berry phase \cite{pancharatnam1956,berry} for discrete paths
\cite{vanderbilt}, $\gamma(C)=-{\rm Im}\ln[ \la \phi_1|\phi_2\ra \la
  \phi_2|\phi_3\ra \cdots \la \phi_{N-1}|\phi_N\ra]$. Here
$\{\phi_i\}$ are $N$ number of nonorthogonal states and $C$ refers to
the close path in the functional space.  We call
$\boldsymbol{\eta}^{(3)}(C)$ the third-order Pancharatnam-Berry
momentum tensor (PB tensor).  It obeys the sum rule as \be
\sum_{mf}(E_i-E_m)(E_m-E_f)(E_f-E_i)\boldsymbol{\eta}^{(3)}(C)=\la
i|[H,\bp]^3|i\ra, \ee which is zero for a system with inversion
symmetry.  This connects the experimental detectables to those
undetectable, which forms the basis of our study \cite{sm}.

\section{Results}

\begin{figure}
%\clr
  %\includegraphics[angle=0,width=\columnwidth]{/home/gpzhang/doe/paper/mn2ruga/weyl/ob/mrg5dos1/dos.eps}
%\includegraphics[angle=0,width=\columnwidth]{/home/gpzhang/doe/paper/mn2ruga/PRB2022/figures/dirac/mrg3/band/dec.3/bandmrg3.eps}
\includegraphics[angle=0,width=0.6\columnwidth]{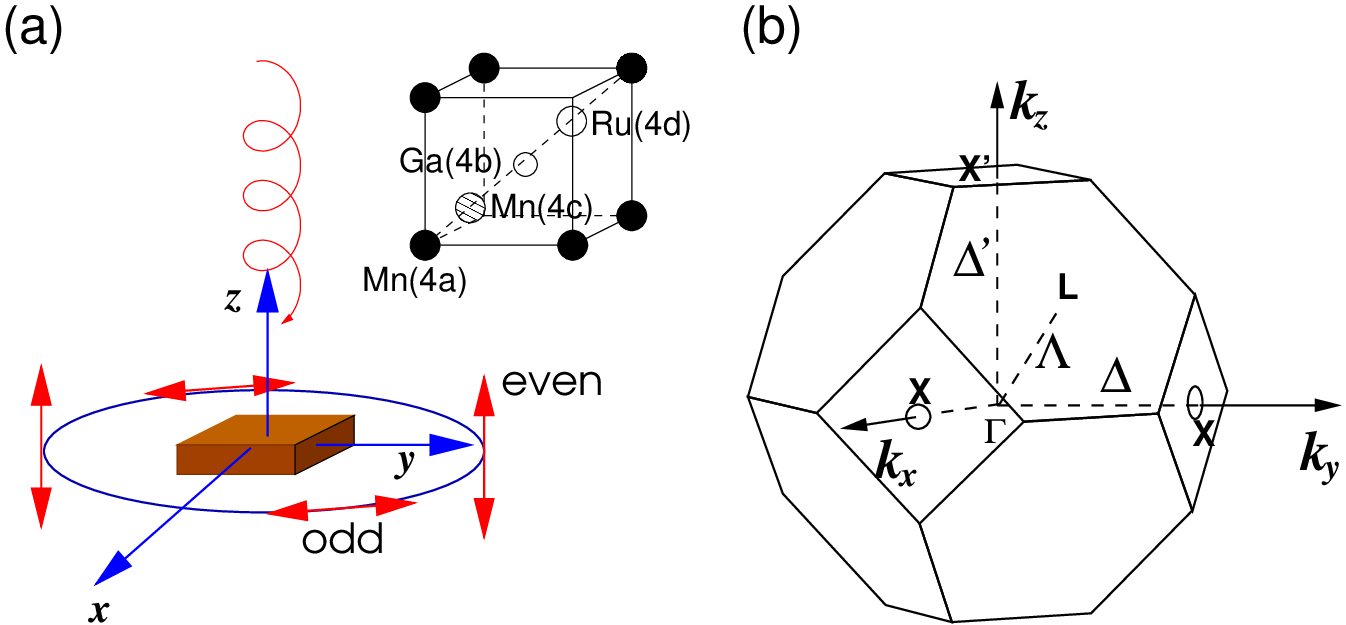}

\includegraphics[angle=0,width=0.6\columnwidth]{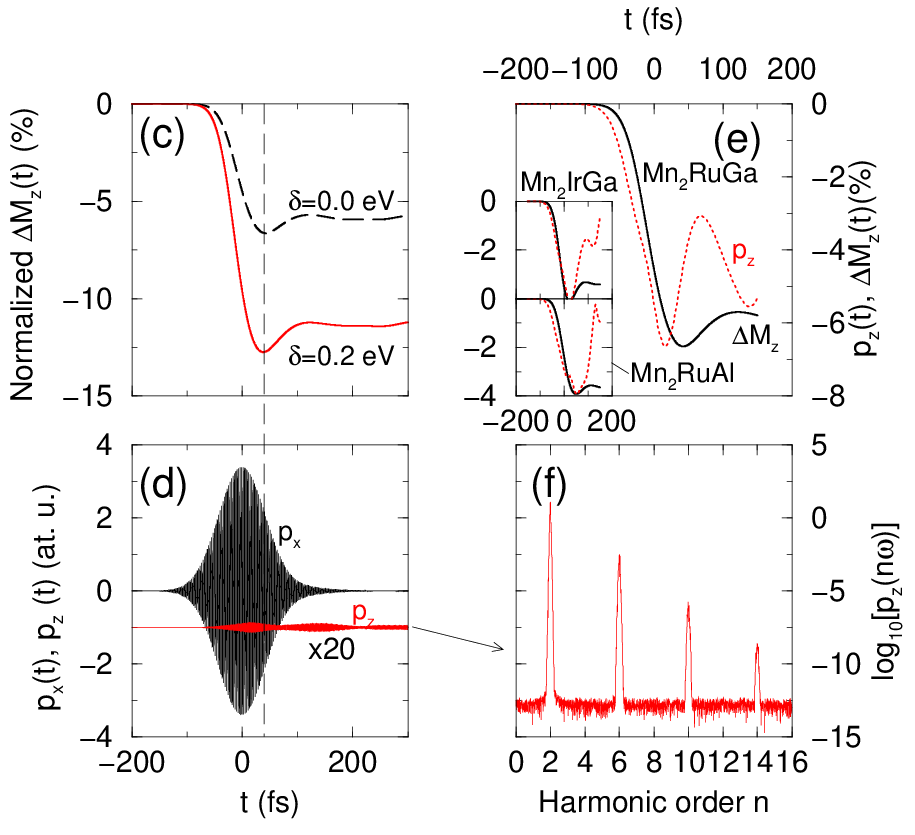}
\caption{(a) Schematic of the interaction between a $\sigma$ pulse and
  \mne.  Even and odd harmonics are emitted along two different
  directions. Inset: Simplified crystal structure of \mne, where
  the filled circles are Mn$_{4a}$, and the shaded circle is Mn$_{4c}$,
  the small and large empty circles are the Ga$_{4b}$ and Ru$_{4d}$
  atoms, respectively.  (b) The Brillouin zone of \mne. (c) Spin
  change $\Delta M_z=(\frac{M_z(t)}{M_z(-\infty)}-1)$ for two
  intraband cutoffs $\delta=0.0$ (dashed line) and 0.2 eV (solid
  line).  (d) $p_x$ and $p_z$ as a function of time.  (e) Envelope
  of $p_z$ (red dotted lines) and $\Delta M_z$ (solid lines) changes
  with time for \mne, $\rm Mn_2IrGa$ and $\rm Mn_2RuAl$ (insets).  (f)
  Logarithmic harmonic signals along the $z$ axis.
% Computed at mrg5dos1.
 }
\label{fig1}
\end{figure}

\mn has a space group symmetry of $\rm F\bar{4}3m$,
with no inversion symmetry.  Figure \ref{fig1}(a) is a simplified
version of the structure, where one Mn is situated at the $4a$ site
(filled circle)  and another Mn at the $4c$ site (shaded circles). Ru
and Ga take the $4d$ and $4b$ sites (small and large empty circles),
respectively.
%$a$ is 5.97 \AA, which matches
%GaAs of 5.75 \AA, within 3.6 \%,
Without SOC, only one term $\la p
\ra^{(2)}_{3,1,2}(t)$ is independent, where $(312)$ are cyclic
Cartesian indices. With SOC, there are three, and we choose $\la p
\ra^{(2)}_{3,1,2}(t)$, so even-order signals polarize along the $z$
axis and are spatially separated from odd orders whose polarization is
along the $x$ and $y$ axes.  Figure \ref{fig1}(a) illustrates this
process, where the light propagates along the $-z$ axis with its
vector potential in the $xy$ plane, ${\bf
  A}(t)=A_0\exp(-t^2/\tau^2)(\cos\omega t \hat{x} \pm \sin\omega t
\hat{y} )$.  Here $A_0$ is the amplitude, $\tau$ is the laser pulse
duration, $t$ is the time, and $\hat{x}$ and $\hat{y}$ are the unit
vectors of the $x$ and $y$ axes, respectively.
{\clr In the presence of SOC, we include all the valence states from
  $-0.4124$ Ry up to the Fermi energy of 0.662 Ry, as well as
  the partially and
completely unoccupied states up to 2.156 Ry. In total, there are 76
states at each $\bk$ point. Semicore and core states are not
included, since they appear below  $-2.37$ Ry. Including more states
produces no visible effect.
We use an extremely
dense $\bk$ mesh of $48\times 48\times 48$ for all the results here, while
the test of $\bk$ convergence is given in the SM \cite{sm}.} 
Figure \ref{fig1}(c)
shows ultrafast demagnetization (long dashed line) under a laser pulse
of $\tau=60 $ fs, $\hbar\omega=1.6$ eV and $A_0=0.015\ \rm V
fs/$\AA. So the fluence is 1.33 \mje.  Including the intraband
transition (the red line with $\delta=0.2$ eV) \cite{jpcm23,sm}
bolsters the demagnetization from 7\% to 13\%, consistent with the
experimental findings \cite{bonfiglio2021}. $\Delta M_z(t)$ is very
smooth without rapid oscillation. By contrast, the charge response,
i.e., $p_x(t)$ and $p_z(t)$, beats strongly.  Figure \ref{fig1}(d)
illustrates that $p_x(t)$ closely follows the laser field and peaks at
0 fs, which is ahead of the spin by 40 fs.  The vertical line across
(c) and (d) highlights this delay of the spin with respect to the
charge response \cite{prb98}.  $p_z(t)$ is much weaker and does not
peak at 0 fs. We magnify $p_z$ by 20 times and then vertically shift
it down by one unit. Since $p_z$ still has rapid oscillations, we take
its negative envelope and superimpose it on the spin change in
Fig. \ref{fig1}(e) (see the solid line). We find that the leading edge
of $p_z(t)$ (dashed line) matches that of $\Delta M_z(t)$, far better
than $p_x(t)$.  We have tested three additional Heuslers, $\rm
Mn_2RuAl$, $\rm Mn_2IrGa$, and $\rm Mn_2RuGe$.  We find that only $\rm
Mn_2RuGe$ has a stronger deviation (see \cite{sm}), and the rest show
a consistent match between $p_z(t)$ and $\Delta M_z(t)$ (see two
insets for $\rm Mn_2RuAl$ and $\rm Mn_2IrGa$).
%testing data are in  /home/gpzhang/doe/paper/mn2ruga/hhg/data/qn/Al1
%testing data are in
%/home/gpzhang/doe/paper/mn2ruga/hhg/data/ob/ir1/rc1.6ev
This demonstrates that our numerical result confirms our above
analysis.  This match is important as it allows us to
use $p_z(t)$ to understand $\Delta M_z(t)$.  Obviously, a perfect
match is not expected.

\begin{figure}
%\clr
  \includegraphics[angle=0,width=0.9\columnwidth]{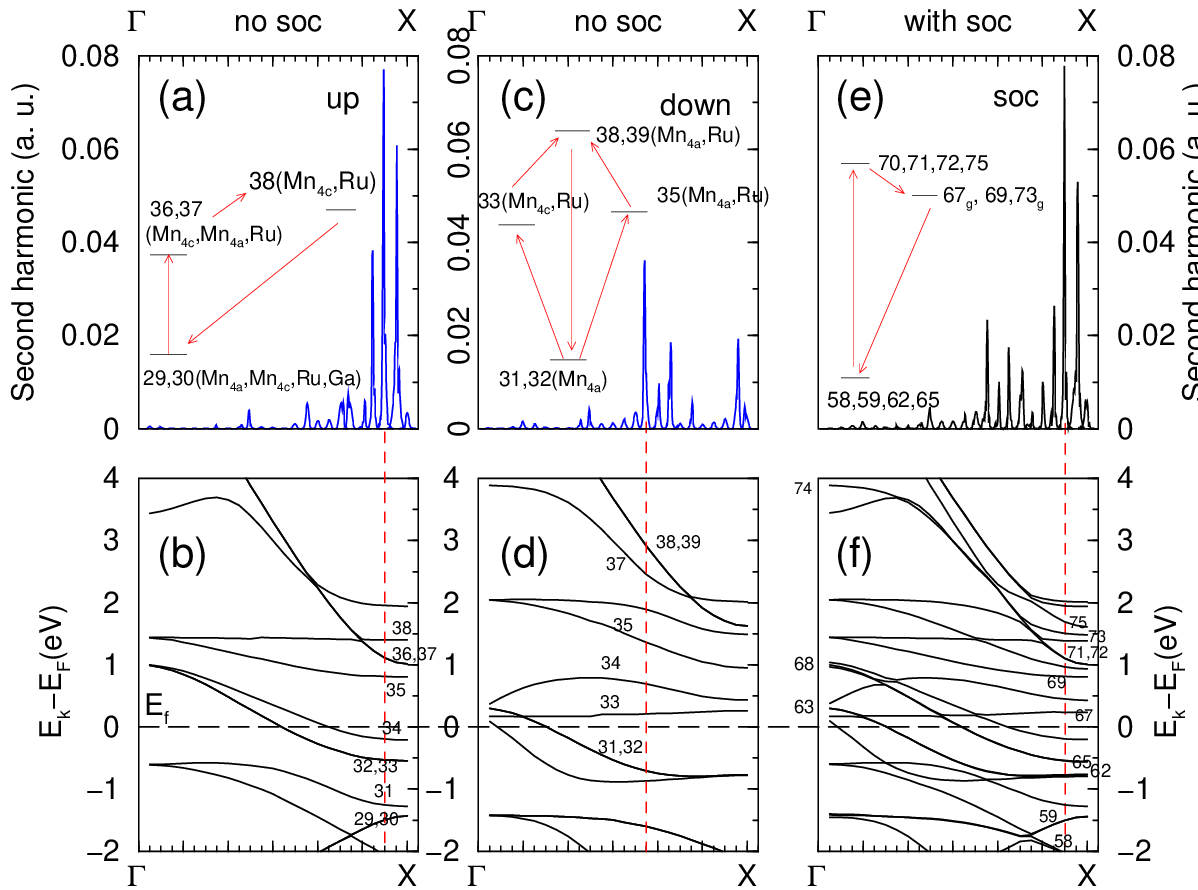}
\caption{ (a) Second harmonic signal of the spin majority channel
  dispersed along the $\Gamma$-X direction, without spin-orbit
  coupling.  The vertical one denotes the location of the largest peak
  at {\clr $(\frac{44}{48},0,0)\frac{2\pi}{a}$}.  (b) Band
  structure of the spin majority channel, where the horizontal dashed
  line is the Fermi level.  (c) The same as (a), but for the spin
  minority channel, where the maximum peak is shifted to {\clr
  $(\frac{30}{48},0,0)\frac{2\pi}{a}$}.  (d) Spin
  minority band structure. (e) Second harmonic signal with
  spin-orbit coupling, whose maximum peak is shifted back to {\clr 
  $(\frac{44}{48},0,0)\frac{2\pi}{a}$}. The numbers with
  subscript $g$ refer to the flat band gateway states.  (f) Band
  structure with SOC. }
\label{fig2}
\end{figure}

We Fourier transform $p_z(t)$ into the frequency domain
(Fig. \ref{fig1}(f)), where the second order dominates, and then
disperse it along the $\Gamma$-X direction (Fig. \ref{fig1}(b)).
Figure \ref{fig2}(a) shows such a plot for the spin majority channel
under $\sigma^+$ excitation. We notice that most $\bk$ points along
the $\Gamma$-X direction have a weak harmonic signal. The strongest
one is at {\clr$(\frac{44}{48},0,0)\frac{2\pi}{a}$}, close to the X
point, {\clr where $a$ is the lattice constant}.  In
Fig. \ref{fig2}(b), we show the band structure for the majority band,
where we draw a dashed line vertically across Figs. \ref{fig2}(a) and
\ref{fig2}(b) to pinpoint which bands are involved in harmonic
generation.  {\clr 
  Our
  Fermi energy without SOC is at $E_F=0.652$ Ry. We number our band states from
  Ga's $3p$ states at $E_{3p}(\rm Ga)=-6.252$ Ry, i.e., from 1 to
  3. States 4 and 5 are Mn's $3s$ states at $E_{3s}({\rm
    Mn}_{4c})=-5.288$ Ry and $E_{3s}({\rm Mn}_{4a})=-5.08$ Ry. State 6
  is Ru's $5s$ state at $E_{5s}({\rm Ru})=-4.67$ Ry, followed by
  six Mn's $3p$ states from 7 to 12 at $E_{3p}({\rm Mn}_{4c})= -2.990$
  and $E_{3p}({\rm Mn}_{4a})=-2.466$ Ry, respectively. Finally, states
  13 to 15 are Ru's $5p$ state at $E_{5p}({\rm Ru})=-2.466$
  Ry. The valence bands start from band 16 at $-0.403$ Ry.  Figure
  \ref{fig2}(b) only shows bands from \ket{29} to \ket{38}. 
A particular challenge in \mrg is that bands are
  strongly hybridized, so it is rather cumbersome to denote them by
  their orbital characters. 
  Table \ref{table2} shows that 
  \ket{29} and \ket{30} are a hybridized band formed by
  Mn$_{4a}$-$3d$, Mn$_{4c}$-$3d$, Ru-$4d$ and Ga-$4p$ states, while
  \ket{38} is dominated by Mn$_{4c}$-$3d$ and Ru-$4d$. A detailed
discussion is given in Table II of SM \cite{sm}.}  Figure \ref{fig2}(c) shows
the second harmonic from the spin minority channel.  Because the
minority channel has nearly zero density of states (DOS) at the Fermi
level \cite{prb22,pra23}, this cuts off signals significantly. Its
maximum is only half the majority channel and is shifted to a
different point, {\clr $(\frac{30}{48}, 0,0)\frac{2\pi}{a}$.}

\begin{figure}
%\clr
  %  \includegraphics[angle=0,width=0.6\columnwidth]{/home/gpzhang/doe/paper/mn2ruga/hhg/data/cori/mn2rugea4/moment/ge.eps}
%  \includegraphics[angle=0,width=1\columnwidth]{/home/gpzhang/doe/paper/mn2ruga/hhg/data/ob/mrg8G3a/moment/g-l.eps}
 \includegraphics[angle=0,width=0.6\columnwidth]{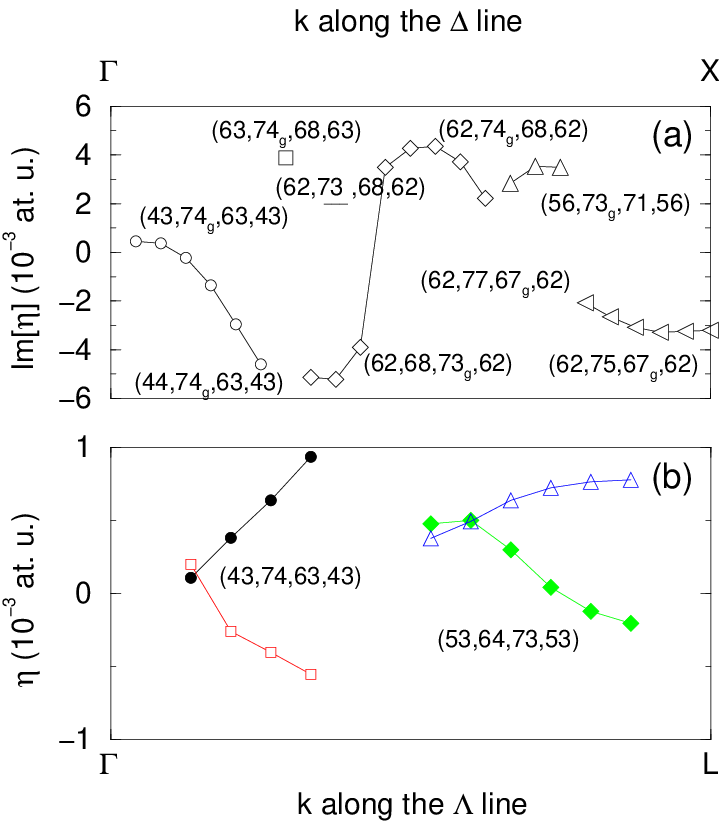}
 \caption{(a) \clr Large elements of the Pancharatnam-Berry tensor
   $\eta^{(3)}(\bk)$ in units of $\hbar^3/a_0^3$ along the $\Gamma$-X
   direction for a few selected channels denoted by ($i,m,f,i$) in the
   figure, same as \chl{i}{m}{f}{i} in the main text, where band
   indices $i,m,f$ are those bands shown in Fig. \ref{fig2}(f). We
   link symbols by lines only if they have similar channel states. For
   instance, the circles are mainly due to \chl{43}{74}{63}{43}.  All
   the elements are all imaginary.  (b) Pancharatnam-Berry tensor
   $\eta^{(3)}(\bk)$ elements along the $\Gamma$-L direction, which is
   much weaker.  The filled symbols are the real part of
   $\eta^{(3)}(\bk)$, while the unfilled symbols are the imaginary
   part of $\eta^{(3)}(\bk)$.  }
\label{fig3}
\end{figure}

{\clr The PB tensor $\eta^{(3)}(\bk)$ has three band indices, $i$, $m$
  and $f$, for each transition channel, written as \chl{i}{m}{f}{i}.
  Most channels have a small element.  We disperse the largest
  elements along the $\Gamma$-X direction ($\Delta$ line) in
  Fig. \ref{fig3}(a).  At $\Gamma$, we find the first dominant
  channel is \chl{43}{74}{63}{43} (see the circles), where bands
  \ket{74} and \ket{63} are shown in Fig. \ref{fig2}(f) but \ket{43}
  is not because its energy is too low.   As we move away from
  $\Gamma$, channel 
  \chl{43}{74}{63}{43} 
  smoothly transitions to \chl{44}{74}{63}{43}
  Once we encounter a band crossing, $\eta^{(3)}(\bk)$
  suddenly changes (see the squares and diamonds).}  Whenever there is
no crossing, a smooth change is found from one $\bk$ to another. This
further confirms that the PB tensor is well defined \cite{sm}. A much
smaller tensor element is found along the $\Gamma$-L line ($\Lambda$
line), because two flat bands are absent along this direction
\cite{galanakis2002a,galanakis2002b,prb22}, where $\eta^{(3)}(\bk)$ is
below our numerical uncertainty limit.  Because of multiple band
crossings, we only find two regions with a smooth dispersion. One is
between {\clr $(\frac{1}{48}, \frac{1}{48}, -\frac{1}{48})\ib$ and
  $(\frac{5}{48}, \frac{5}{48}, -\frac{5}{48})\ib$, and the other is
  between $(\frac{8}{48}, \frac{8}{48}, -\frac{8}{48})\ib$ and
  $(\frac{13}{48}, \frac{13}{48}, -\frac{13}{48})\ib$.} Different from
the $\Gamma$-X line, $\eta^{(3)}(\bk)$ has both real (filled symbols)
and imaginary parts (empty symbols).  In summary, we have six
channels: four for the minority, ${|31\ra}
\substack{\nearrow\\\searrow} \begin{matrix}|33\ra\\|35\ra
\end{matrix}
\begin{matrix}\rightarrow\\\rightarrow
\end{matrix}
\begin{matrix}|38\ra\\|38\ra
\end{matrix}
\substack{\searrow\\\nearrow}
{|31\ra}, 
{|32\ra} \substack{\nearrow\\\searrow} \begin{matrix}|33\ra\\|35\ra
\end{matrix}
\begin{matrix}\rightarrow\\\rightarrow
\end{matrix}
\begin{matrix}|39\ra\\|39\ra
\end{matrix}
\substack{\searrow\\\nearrow}
|{32}\ra, $
{\clr where the band indices are shown in Fig. \ref{fig2}(d),}
and two for the
majority, $
|{29}\ra\rightarrow
|{36}\ra \rightarrow |{38}\ra \rightarrow |{29}\ra,
|{30}\ra\rightarrow |{37}\ra
\rightarrow |{38}\ra \rightarrow |{30}\ra, 
$ {\clr where indices are shown in Fig. \ref{fig2}(b).}
\iffalse
\begin{widetext}
\ba
{|31\ra} \substack{\nearrow\\\searrow} \begin{matrix}|33\ra\\|35\ra
\end{matrix}
\begin{matrix}\rightarrow\\\rightarrow
\end{matrix}
\begin{matrix}|38\ra\\|38\ra
\end{matrix}
\substack{\searrow\\\nearrow}
{|31\ra}, &
{|32\ra} \substack{\nearrow\\\searrow} \begin{matrix}|33\ra\\|35\ra
\end{matrix}
\begin{matrix}\rightarrow\\\rightarrow
\end{matrix}
\begin{matrix}|39\ra\\|39\ra
\end{matrix}
\substack{\searrow\\\nearrow}
|{32}\ra,&  
 {\rm (Minority)}\\
 |{29}\ra\rightarrow
|{36}\ra \rightarrow |{38}\ra \rightarrow |{29}\ra,&
|{30}\ra\rightarrow |{37}\ra
\rightarrow |{38}\ra \rightarrow |{30}\ra. &            {\rm (Majority)}
\ea
\end{widetext}
\fi Figure \ref{fig2}(b) reveals that the flat band $|38\ra$ is at the
center of the majority channel excitation.  We call it a gateway
state.  The gateway state in the minority channel is \ket{33}, which
is also a flat band (see Fig. \ref{fig2}(d)).  Including SOC does not
change the picture, except that the harmonic maximum shifts back to
%$\bk_{593}$
{\clr$(\frac{44}{48},0,0)\frac{2\pi}{a}$}
(Fig. \ref{fig2}(e)) and the flat bands are renumbered due
to the doubling number of bands (Fig. \ref{fig2}(f)). They are now
\ket{67} and \ket{73}.

\begin{figure}
%\clr
  %  \includegraphics[angle=0,width=0.6\columnwidth]{/home/gpzhang/doe/paper/mn2ruga/hhg/data/cori/mn2rugea4/moment/ge.eps}
    \includegraphics[angle=0,width=1\columnwidth]{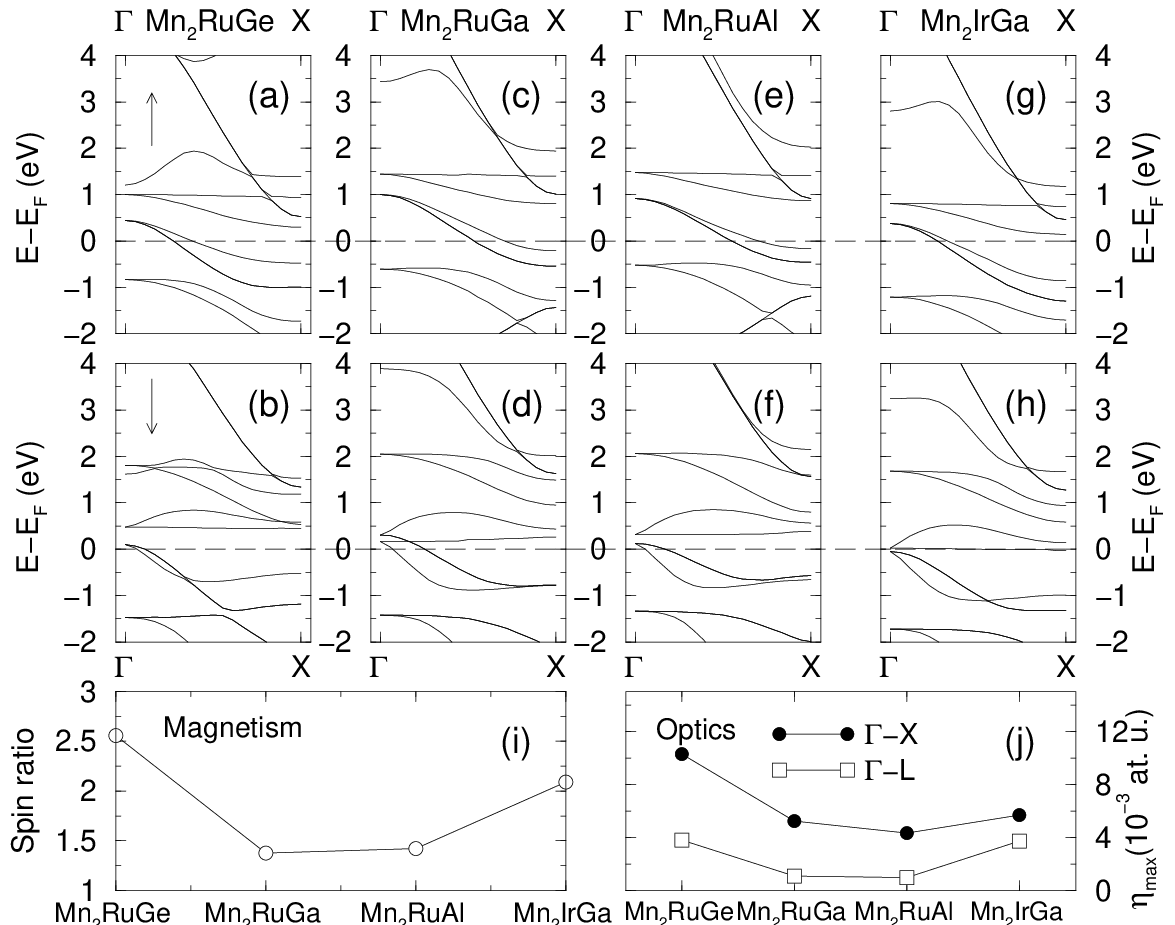}
  \caption{(a)-(h) Comprehensive band structure comparison among four
    Heusler compounds, $\rm Mn_2RuGe$, $\rm Mn_2RuGa$, $\rm Mn_2RuAl$
    and $\rm Mn_2IrGa$. The majority bands are in (a), (c), (e) and
    (g), while the minority are in (b), (d), (f) and (h).
    \if
    (a) Spin majority band of Mn$_2$RuGe. Mn$_{4a}$ has spin
    moment of 2.94913 $\mu_B$, while Mn$_{4c}$ has $-1.1519\mu_B$. Ru
    and Ge have spin moments of 0.12011 and 0.003491 $\mu_B$. 
The ratio is 2.5602.
    (b) Spin majority
    band of Mn$_2$RuGa.
 Mn$_{4a}$ has spin
    moment of 3.17551  $\mu_B$, while Mn$_{4c}$ has $-2.30834\mu_B$. Ru
    and Ga have spin moments of 0.07641  and 0.03372 $\mu_B$.  The
    ratio is 1.3757.
    (c)
    In Mn$_2$IrGa, Mn$_{4a}$ is  3.20507, Mn$_{4c}$ is  -1.53271
    Ir is 0.23870, Ga is 0.02517 $\mu_B$. The ratio is 2.091113126.
    (d) In Mn$_2$RuAl,  Mn$_{4a}$ is  3.07633, Mn$_{4c}$ is $-2.16601$
    Ru is 0.05448, Al is 0.01907 $\mu_B$. The ratio is 1.420275068.
    \fi
    (i) Spin moment ratio between two Mn atoms. (j) Maximum PB tensor
    elements $\eta^{(3)}(\bk)$ along the $\Gamma$-X and $\Gamma$-L directions. 
  }
\label{fig4}
\end{figure}

Are these flat bands essential to AOS? We employ three additional
Heuslers, $\rm Mn_2YZ$, where Y is Ir \cite{patel2023} and $Z$ is Ge
and Al \cite{endo2012,roy2021}.  Figures \ref{fig4}(a) and
\ref{fig4}(b) show that Mn$_2$RuGe has two similar flat bands, but its
Fermi level is raised clearly with respect to \mn (compare with
Figs. \ref{fig4}(c) and \ref{fig4}(d)).  Replacing Ga by Al has the
smallest effect on the band structure among all three new Heuslers
(see Figs. \ref{fig4}(e) and \ref{fig4}(f)). The flat bands are
retained. When we replace Ru by Ir \cite{patel2023}, the flat band in
the majority is lowered, and that in the minority is now at the Fermi
level, which explains why it has a giant coercivity \cite{patel2023}.

We recall that in GdFeCo and TbFe the spin moment ratios of two spin
sublattices are always very close to each other \cite{epl16}. We
compute the ratio between the spin moment on $\rm Mn_{4a}$ and $\rm
Mn_{4c}$. The results are shown in Fig. \ref{fig4}(i). To our
surprise, there is a significant variation.  $\rm Mn_2RuGa$ and $\rm
Mn_2RuAl$ have the smallest ratio, while $\rm Mn_2RuGe$ and $\rm
Mn_2IrGa$ have a larger ratio.  The smaller ratio between two spin
sublattices is consistent with the small remanence criteria
\cite{hassdenteufel2015} proposed much earlier.  Figure \ref{fig4}(j)
reveals the optical requirement of AOS through the PB tensor
$\eta^{(3)}(\bk)$. We see that $\rm Mn_2RuGe$ has the strongest value
and \mn has the smaller value. Because a strong optical excitation and
a strong magnetic excitation are always opposite to each other, it is
likely that the competition between these two underlines the
all-optical spin switching.

\section{Connection to future experiments}

An experimental test of our theoretical prediction is exactly what we
had in our mind when we carried out this study.  For this reason, all
the Heusler compounds, Mn$_2$RuAl, Mn$_2$RuSn, Mn$_2$RuSi, Mn$_2$RuGa,
Mn$_2$IrGa and Mn$_2$RuGe, are already synthesized experimentally or
well studied theoretically. The lattice constant $a$ of Mn$_2$RuGa,
5.97 \AA, matches that of GaAs, 5.75 \AA, within 3.6 \%,
very attractive integration of a semimetal into a state-of-the-art
semiconductor, to extend the boundary of spintronics.

We have three suggestions for experimentalists. First, since
Mn$_2$RuAl has similar electronic, magnetic and optical properties as
Mn$_2$RuGa, we would expect that the chance to observe AOS in
Mn$_2$RuAl is high. So an experimental test of AOS on Mn$_2$RuAl will
test our first prediction.

Second, Mn$_2$RuGe differs from Mn$_2$RuGa
by an extra electron in the outer shell.  Mn$_2$RuGe has the largest
spin ratio among all four Heuslers investigated. We do not expect a
single pulse AOS in this system. Whether or not AOS occurs in
Mn$_2$RuGe will directly test our magnetic moment ratio prediction.

Third, Mn$_2$IrGa presents another opportunity.  Here Ir (in
Mn$_2$IrGa) and Gd (in GdFeCo) both have $4f$ and $5d$ shells.
Although Gd is different from Ru and Ir, Gd is the only one among $4f$
rare-earth elements that has $5d$ electrons and is thus similar to Ru
and Ir, whereas $4f$ electrons play a minor role as demonstrated in
magnetic anisotropy \cite{abdelouahed2009}.  Ir's $4f$ shell is completely
filled, since its native electron configuration is
Xe$4f^{14}5d^76s^2$.  $4f$ in Gd is half filled since its
configuration is Xe$4f^{7}5d6s^2$.  They both have zero orbital moment
from the $4f$ shell. In addition, Ir also has zero spin moment from
the $4f$ shell. Whether AOS in Mn$_2$IrGa occurs experimentally allows
one to distinguish the role of $4f$, with respect to their differences
in $5d$. Since Ru's native electron configuration is Kr$4d^75s^1$, a
comparison between Mn$_2$IrGa and Mn$_2$RuGa is potentially able to
compare and contrast the roles of $4d$ and $5d$ in AOS. Therefore, we
believe there are ample opportunities that await experimental
verifications.

\iffalse

The ratio is 2.5602.  (b) Spin majority band of Mn$_2$RuGa.  Mn$_{4a}$
has spin moment of 3.17551 $\mu_B$, while Mn$_{4c}$ has
$-2.30834\mu_B$. Ru and Ga have spin moments of 0.07641 and 0.03372
$\mu_B$.  The ratio is 1.3757.  (c) In Mn$_2$IrGa, Mn$_{4a}$ is
3.20507, Mn$_{4c}$ is -1.53271 Ir is 0.23870, Ga is 0.02517
$\mu_B$. The ratio is 2.091113126.  (d) In Mn$_2$RuAl, Mn$_{4a}$ is
3.07633, Mn$_{4c}$ is $-2.16601$ Ru is 0.05448, Al is 0.01907
$\mu_B$. The ratio is 1.420275068.

\fi

\iffalse

WAIT A particular interesting case is found, when we work on Mn2RuGe,
    where we have replaced Ga by Ge. In this case, we find that the
    product in the spin majority channel is very small, but the spin
    minority channel is very large. By contrast, in \mne, we see these
    two channels are comparable. This is the main difference that is
    central to AOS should be tested experimentally. The band structure
    of both compounds are similar to some extent. It is important to
    point out their ground state spin states are different. The
    magnetic spin moments

Ir: converges very quickly from k16 to k36. the leading edge
in particular

Ga: is similar to ir

Al: has a larger effect with number of k point. With more k points,
the spin reduction is more.

Ge: converges the fastest. The entire leading edge is the same.

\fi

\section{Conclusion}

We have shown that a group of flat bands, with a large
PB tensor, act as gateway states that channel one sublattice spin to
another to realize all-optical spin switching in Mn-based Heusler
ferrimagnets.  This is the origin of the theoretical intersite
exchange model by Mentink \et \cite{mentink2012}.  Although all four
Heuslers have flat bands, only \mn and $\rm Mn_2RuAl$ have a small
spin moment ratio between two spin sublattices, explaining the
experimental low remanence criterion \cite{hassdenteufel2015}. Our
finding will have a significant impact on future research on two
fronts.  Conceptually, the Pancharatnam-Berry tensor extends the Berry
phase \cite{berry} and connects the $n$th-order nonlinear optics to
the $(n+1)$th PB tensor, so one has a way to quantify the efficacy of
nonlinear optical materials. More importantly, since all four Heuslers
are readily available, their supertunability renders them an ideal
test bed, representing a paradigm shift for future experimental and
theoretical investigations of all-optical spin switching.

\acknowledgments We would like to thank Dr. Peter Blaha for helpful
communications on the partial charge. YQL and MSS were supported by
the National Science Foundation of China under Grant No. 11874189.
GPZ was partly supported by the U.S. Department of Energy under
Contract No.  DE-FG02-06ER46304.  Part of the work was done on Indiana
State University's Quantum and Obsidian Clusters.  This research used
resources of the National Energy Research Scientific Computing Center,
which is supported by the Office of Science of the U.S.  Department of
Energy under Contract No. DE-AC02-05CH11231.

\iffalse $\begin{pmatrix} 0 &-1 & 0\\ 1 &0&0 \\ 0&0&-1 \\
    \end{pmatrix}$\fi

%{Availability of data.} The data that support the findings of this
%study are available from the corresponding author upon reasonable
%request.

$^*$guo-ping.zhang@outlook.com.
https://orcid.org/0000-0002-1792-2701

%\begin{widetext}

    \begin{table}
      \centering
        \caption{\clr (Top) Composition of the majority band states at
          $(\frac{44}{48},0,0)\frac{2\pi}{a}$ and (bottom) that of
         the minority band states at
          $(\frac{30}{48},0,0)\frac{2\pi}{a}$. {\clr The band indices are
         shown in Fig. \ref{fig2}.} 
          $l$ refers to the orbital angular momentum quantum
          number. The numbers given under each column are partial
          charges in 
          percentages. Since the majority and minority states are from
        different $\bk$, one should not add them up. }
        \label{table2}
        \begin{tabular}{c|c|c|c|c|c|c|c|c|c|c|c|c}
\hline
\hline
       Majority  states& \multicolumn{3}{|c|}{Mn$_{4a}$}&
          \multicolumn{3}{|c|}{Mn$_{4c}$}&  
          \multicolumn{3}{|c|}{Ru} &
          \multicolumn{3}{|c}{Ga}\\
\hline
{\clr  $(\frac{44}{48},0,0)\frac{2\pi}{a}$}
& $l=0$& $l=1$& $l=2$& $l=0$& $l=1$& $l=2$& $l=0$& $l=1$& $l=2$ &
$l=0$& $l=1$& $l=2$ \\
\hline
\ket{29},\ket{30} & 0 & 3.71 & 13.88& 0& 5.78 & 8.92 &0& 6.41 & 10.57 &0& 18.25&
0.01\\
%30 & 0 & 3.71 & 13.88& 0& 5.78 & 8.92 &0& 6.41 & 10.57 &0& 18.25&
%0.01\\
%35 & 0.39 & 0 & 0.53& 0.03& 0.02 & 62.33 &0& 0.02 & 31.44 &0.32& 0.01&
%0.04\\
\ket{36},\ket{37} & 0    & 1.15 & 18.46 & 0& 0.82 & 59.52 &0& 0.98 & 10.48 &0& 0.57&
1.42\\
%37 & 0    & 1.15 & 18.46 & 0& 0.82 & 59.52 &0& 0.98 & 10.48 &0& 0.57&
%1.42\\
\ket{38} & 0    & 0    & 0     & 0& 0    & 71.85 &0& 0    &  25.84& 0& 0&0\\
        \end{tabular}
             \begin{tabular}{c|c|c|c|c|c|c|c|c|c|c|c|c}
\hline
\hline
       Minority   states& \multicolumn{3}{|c|}{Mn$_{4a}$}&
          \multicolumn{3}{|c|}{Mn$_{4c}$}&  
          \multicolumn{3}{|c|}{Ru} &
          \multicolumn{3}{|c}{Ga}\\
\hline
{\clr $(\frac{30}{48},0,0)\frac{2\pi}{a}$}
& $l=0$& $l=1$& $l=2$& $l=0$& $l=1$& $l=2$& $l=0$& $l=1$& $l=2$ &
$l=0$& $l=1$& $l=2$ \\
\hline
\ket{31},\ket{32} & 0 & 2.50 & 45.29& 0& 8.33 & 0.67 &0& 5.72 &  3.78 &0&  7.68&
0.58\\
%32 & 0 & 2.50 & 45.29& 0& 8.33 & 0.67 &0& 5.72 &  3.78 &0&  7.68&
%0.58\\
\ket{33} & 0    & 0 & 4.57& 0   & 0    & 51.60 &0& 0    & 40.00 &0   & 0   &
0.15\\
%34 & 0.05 & 0.63 & 44.78 & 0.23& 0    &  39.24 &3.64& 0.95 & 2.05 &0.51& 1.21&
%0.26\\
\ket{35} & 0    & 0    & 78.75 & 0& 0    & 0.01  &0& 0    & 11.45 &0& 0   &
0.69\\
%37 & 0.22    & 3.35    & 46.13     & 4.79&1.37    & 1.67 &2.60&1.02  &  9.29& 10.89& 4.17&0.10\\
\ket{38},\ket{39} & 0    & 3.87    & 42.54     & 0& 3.92    & 8.67 &0& 3.37    & 16.46& 0& 4.51&1.31\\
%39 & 0    & 3.87    & 42.54     & 0& 3.92    & 8.67 &0& 3.37    & 16.46& 0& 4.51&1.31\\
\hline\hline
        \end{tabular}   
    \end{table}
%\end{widetext}

%\onecolumngrid

%  \includegraphics[angle=0,width=0.8\columnwidth]{/home/gpzhang/doe/paper/mn2ruga/hhg/data/qn/mrgx3/g-x/band.eps}

\end{document}